\documentclass[lettersize,journal]{IEEEtran}
\usepackage{amsmath,amsfonts}
\usepackage{algorithmic}
\usepackage{algorithm}
\usepackage{array}
\usepackage[caption=false,font=normalsize,labelfont=sf,textfont=sf]{subfig}
\usepackage{textcomp}
\usepackage{stfloats}
\usepackage{url}
\usepackage{verbatim}
\usepackage{graphicx}
\usepackage{cite}
\usepackage{hyperref} 
\usepackage{tikz,xcolor}
\usepackage{orcidlink}
\usepackage{multirow}
\usepackage{booktabs} 
\hypersetup{hidelinks}
\hypersetup{
colorlinks=true,
linkcolor=blue,
citecolor=blue
}
\definecolor{lime}{HTML}{A6CE39}
\DeclareRobustCommand{\orcidicon}{%
    \begin{tikzpicture}
    \draw[lime, fill=lime] (0,0) 
    circle [radius=0.16] 
    node[white] {{\fontfamily{qag}\selectfont \tiny ID}};    \draw[white, fill=white] (-0.0625,0.095) 
    circle [radius=0.007];    \end{tikzpicture}
    \hspace{-2mm}}
\foreach \x in {A, ..., Z}{%
    \expandafter\xdef\csname orcid\x\endcsname{\noexpand\href{https://orcid.org/\csname orcidauthor\x\endcsname}{\noexpand\orcidicon}}
    }

\begin{document}

\title{Human Detection in Realistic Through-the-Wall Environments using Raw Radar ADC Data and Parametric Neural Networks}

\author{Wei Wang\orcidA{}, Naike Du, Yuchao Guo, Chao Sun, Jingyang Liu, Rencheng Song\orcidC{},~\IEEEmembership{Member,~IEEE}, 

Xiuzhu Ye\orcidB{},~\IEEEmembership{Senior Member,~IEEE}
\thanks{This work was supported by the National Natural Science Foundation of China under Grant 61971036, the Fundamental Research Funds for the Central Universities under Grant 2023CX01011, and Beijing Nova Program under Grant 20230484361.

Wei Wang, Naike Du, Yuchao Guo, Chao Sun, Jingyang Liu and Xiuzhu Ye are with the School of Information and Electronics, Beijing Institute of Technology, Beijing 100081, China(e-mail: xiuzhuye@outlook.com).

Rencheng Song is with the Department of Biomedical Engineering, and also with the Anhui Province Key Laboratory of Measuring Theory and Precision Instrument, Hefei University of Technology, Hefei 230009, China.}
\thanks{Manuscript received April 19, 2021; revised August 16, 2021.}}

\markboth{Journal of \LaTeX\ Class Files,~Vol.~14, No.~8, August~2021}%
{Shell \MakeLowercase{\textit{et al.}}: A Sample Article Using IEEEtran.cls for IEEE Journals}

\IEEEpubid{\begin{minipage}{\textwidth}\ \\[30pt] \centering
		Copyright \copyright 20xx IEEE. Personal use of this material is permitted. 
		However, permission to use this material for any other purposes must \\ be obtained 
		from the IEEE by sending an email to pubs-permissions@ieee.org.
\end{minipage}}

\maketitle

\begin{abstract}
The radar signal processing algorithm is one of the core components in through-wall radar human detection technology. Traditional algorithms (e.g., DFT and matched filtering) struggle to adaptively handle low signal-to-noise ratio echo signals in challenging and dynamic real-world through-wall application environments, which becomes a major bottleneck in the system. In this paper, we introduce an end-to-end through-wall radar human detection network (TWP-CNN), which takes raw radar Analog-to-Digital Converter (ADC) signals without any preprocessing as input. We replace the conventional radar signal processing flow with the proposed DFT-based adaptive feature extraction (DAFE) module. This module employs learnable parameterized 3D complex convolution layers to extract superior feature representations from ADC signals, which is beyond the limitation of traditional preprocessing methods. Additionally, by embedding phase information from radar data within the network and employing multi-task learning, a more accurate detection is achieved. Finally, due to the absence of through-wall radar datasets containing raw ADC data, we gathered a realistic through-wall (RTW) dataset using our in-house developed through-wall radar system. We trained and validated our proposed method on this dataset to confirm its effectiveness and superiority in real through-wall detection scenarios.
\end{abstract}


\section{Introduction}
\IEEEPARstart{I}{n} recent years, rapid advancements in the field of deep learning and autonomous driving have spurred extensive research into radar-based object detection. Radars exhibit superior robustness and adaptability compared to other sensors in complex and dynamic environments. The capacity to penetrate various weather conditions (e.g., rain and fog), coupled with a broad detection range and the ability to sense object velocities, locations, and sizes, has fueled its widespread usage in various fields, including healthcare, security, and rescue operations. Moreover, compared to sensors such as cameras, radar possesses significant advantages in terms of privacy.

\begin{figure}[!t]
\centering
\includegraphics[width=3.4in]{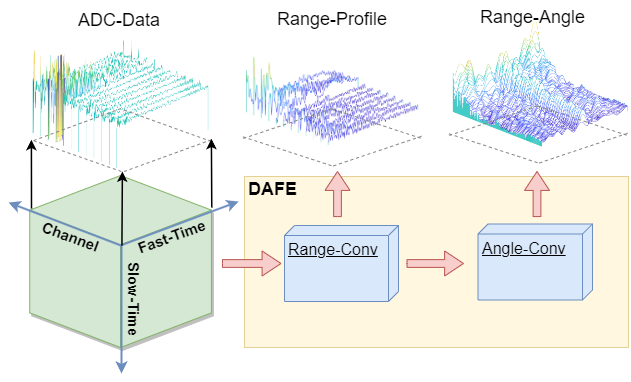}
\caption{Raw ADC data cube and its processing methodology.}
\label{fig_1}
\end{figure}

Through-wall human detection, driven by the manifold advantages and mature development of radar technology, has become a focal point for numerous researchers. Through-wall radar (TWR) utilizes low-frequency ultra-wideband (UWB) signals for their strong penetration ability through obstacles like walls and non-metal barriers. Subsequent to penetrating these obstacles, TWR receives and processes echoed signals generated by human movements, breathing, heartbeat, and other behaviors to detect and track targets. 

Nowadays, numerous works related to through-wall radar have been proposed. Regardless of hardware capabilities, various algorithms for radar signal processing still face significant limitations, making it challenging to effectively detect targets in real through-wall scenarios.Traditional frequency modulated continuous wave (FMCW) radar signal processing entails the initial sampling of the intermediate frequency (IF) signal by the ADC to acquire the raw radar signal. Subsequently, the signal undergoes windowing, filtering, and Discrete Fourier Transform (DFT) operations to extract target distance, velocity, and azimuth information. Finally, Constant False Alarm Rate (CFAR) processing is employed for peak detection, ultimately yielding the target detection results. However, this signal processing approach poses several disadvantages for through-wall tasks: (1) Extensive preprocessing of data may lead to information loss. (2) Due to severe interference of radar signals by walls, the echoed signal experiences significant attenuation along with considerable coupling and noise, resulting in a low signal-to-noise ratio (SNR). DFT, being a predefined equidistant linear transformation, is capable of extracting frequency information from echoed signals (DFT, discussed in Section \ref{dft}). However, the interference signals occupy almost the entire frequency band. At this point, the Discrete Fourier Transform fails to effectively highlight the frequency domain characteristics for extracting through-wall signals and isolating the target. (3) In real through-wall scenarios, the environment is complex and dynamic. Algorithms unable to adaptively process signals hold no practical significance for the application of through-wall radar. 

The through-wall radar algorithm can be divided into two main sections: the signal processing part and the target detection part. Numerous target detection methods based on neural networks have been proposed. They replace the need for manually adjusting parameters and less adaptive CFAR detectors, significantly enhancing the detection performance. However, signal processing, as compared to target detection, is more critical in determining the upper limits of performance for through-wall radar algorithms and the entire system. The lack of further research in this area has become a bottleneck. The methodologies commonly rely on preprocessed radar data. In this paper, we are inspired by works in gesture recognition \cite{zhao2023cubelearn} and autonomous driving \cite{yang2023adcnet} fields, where the raw ADC data from radar is used as input and employ several layers of networks are employed to extract distance, velocity, and azimuth information of targets, replacing the traditional radar data processing pipelines. This approach may effectively addresses issues in through-wall tasks such as difficulties in extracting features from low signal-to-noise ratio data and declining detection performance.

In this work, an end-to-end through-wall human detection neural network is proposed, that aim to utilize raw ADC data to offer comprehensive information while integrating the data processing and feature extraction processes together into the neural network. This integration occurs alongside the training and optimization of the target detector. To the best of the authors' knowledge, this is the first work to utilize 3D convolution in an FMCW radar signal processing pipeline instead of traditional DFT to process raw ADC data for through-wall radar imaging and target detection. The processing method of ADC data is shown in Fig. \ref{fig_1}. It replaces radar signal preprocessing and is integrated into the network. Compared with other target detection networks that use fixed signal processing algorithms to preprocess data, the adaptive learning signal processing algorithm can jump out of the local optimal point in the feature extraction process and achieve higher detection accuracy, especially in challenging datasets with low signal-to-noise ratios. The experiments confirm that this method achieves basic frequency domain feature extraction similar to DFT, while also accomplishing functions like feature selection and noise suppression that DFT cannot perform. 

The main contributions in this work are listed as follows:
\begin{itemize}
\item{We proposed an end-to-end through-wall radar human detection network (TWP-CNN) tailored for real, complex through-wall data.}
\item{The DAFE (DFT-based Adaptive Feature Extraction) module we designed incorporates both DFT and Chirp-Z transform (CZT) parameterization within convolutional layers. Through network learning and optimization, it enables the extraction of subtle target features from ADC data, with noise suppression, and feature data reconstruction.}
\item{We introduced a phase-based channel fusion encoder (PCF-Encoder) along with a coordinate regression module (PCR), which leverage the phase information in radar signals, significantly enhancing the accuracy of target detection and distance-angle position regression.} 
\item{A new dataset, named realistic through-wall (RTW) dataset, which comprises synchronized raw radar ADC data, RGB images, and coordinate labels for the targets is constructed and utilized to prove the validity of the proposed method.}
\end{itemize}

The rest of this paper is organized as follows. Related works for through-wall radar networks and end-to-end networks based on raw ADC data are reviewed in Section \ref{Two}. In Section \ref{Three}, we introduce our through-wall radar system, related algorithms, and proposed network architecture. Then, we describe our self-collected dataset, named RTW, and the experimental design including validation results and visual representations in Section \ref{Four}. Finally, we summarize the entire paper and propose future directions for our work in Section \ref{Five}.

\section{Related Works}
\label{Two}
\subsection{Through-Wall Radar Human Perception}
 Kılıç et al. \cite{kilicc2019through} utilized an SFCW radar to collect 1D range profile, and then employed a CNN for feature extraction, classifying three scenarios: an empty space, standing, and sitting. Nguyen et al. \cite{nguyen2023high} employed a cascaded neural network to estimate the angle of arrival, demonstrating that using deep learning methods can significantly improves the angle resolution of radar target detection. Ding et al. \cite{ding2022human} trained a Convolutional Long Short-Term Memory (ConvLSTM) network to discern human motions from range-Doppler frames. Ding et al. \cite{ding2018application} proposed the utilization of linear prediction coding and data fusion techniques to address the frequency ambiguity issue in target tracking with Doppler through-wall radar. The application of pose reconstruction in through-wall radar imaging is also extensive. Song et al. \cite{song2021efficient} utilizes a 2D antenna array MIMO radar to perform 3D back projection reconstruction on collected through-wall signals. Subsequently, the post-CFAR results are sampled, and a neural network is employed to reconstruct the human body skeleton from the sampled points. Zheng et al. \cite{zheng2021human}, \cite{zheng2023radarformer} utilizes a 3D convolutional neural network for reconstructing the human skeletal structure from thermal images obtained by through-wall imaging. Additionally, they employ a Transformer to directly perceive human presence from radar echo information. This aligns with our shared intuition: the target features embedded within radar echoes can be directly extracted. End-to-end networks excel in extracting information from data, thereby addressing the challenges of through-wall imaging. In previous through-wall human localization efforts, only preprocessed data are utilized for target detection and localization. However, when facing complex and variable scenarios, traditional data preprocessing struggles to achieve the desired suppression of wall-related noise, resulting in a significant performance decline. Li et al. \cite{li2020human} use fully convolutional networks (FCN) to train on simulated through-wall radar data and validate the superiority of the network over various traditional algorithms on experimental data. Pan et al. \cite{pan2023multi} employs an improved version of the U-Net architecture to train on simulated through-wall data. The network incorporates channel attention modules and residual structures. However, these approaches struggle to be effectively applied in practical through-wall scenarios. Wang et al. \cite{wang2023real} utilizes a large-aperture radar to collect radar data in both vertical and horizontal directions. Subsequently, he performs static and dynamic background elimination separately on the input data before feeding it into the network for target detection. This method relies heavily on extensive prior information, and manually conducting background elimination might brought more detrimental than beneficial for low signal-to-noise ratio data. The abundance of information could potentially be lost during the background elimination process. 
Therefore, we attempt to construct an end-to-end network for direct processing of radar's raw ADC data.

\subsection{Learning of Raw ADC Data}
The readability of raw ADC data might be poor, yet it encapsulates all information. Leveraging the powerful fitting capability of neural networks for feature extraction from such data is viable. Ye et al. \cite{ye2019using}, \cite{ye2020human} initially proposed using Short-Time Fourier Transform (STFT) to convert radar ADC data into the frequency domain, which was a hand-crafted representation, not the most optimal. Subsequently, he introduced RadarNet and F-ConvNet, utilizing two one-dimensional convolutional layers instead of STFT for human motion classification. Later, inspired by 1D SincNet \cite{ravanelli2018speaker} in speech processing, Stadelmayer et al. \cite{stadelmayer2023parametric} introduced a network for human motion classification. They integrated a 2D Sinc filter and a 2D Morlet wavelet filter into the neural network, replacing traditional signal processing techniques like windowing and FFT. This led to higher classification accuracy. In the task of gesture recognition, Zhao et al. \cite{zhao2023cubelearn} also proposed the use of multiple linear layers initialized with Discrete Fourier Transform (DFT) parameters to process raw radar data and demonstrated its effectiveness. In \cite{stephan2021radar}, the authors also introduced a DFT parameter layer ahead of the Variational Autoencoder (VAE), using it to reconstruct range-angle image from the obtained RD images. Yang et al. \cite{yang2023adcnet} applied this idea to automotive radar target detection, extracting latent RD information for subsequent network detection. However, in the aforementioned work, radar data was commonly processed by splitting the complex numbers into channels, thus losing crucial phase information present in the radar signal. Our experiments have demonstrated that leveraging phase information can significantly enhance detection accuracy. Moreover, we embedded the Chirp Z-transform (CZT) algorithm (a generalization of the DFT) into the network to tailor it to our task.

\begin{table}[!t]
\caption{Configurations of Radar\label{tab:table1}}
\centering
\begin{tabular}{c   c}
\hline
Configuration & Value\\
\hline
Ramp start frequency $f_{min}$ & 1.8 GHz\\
Ramp end frequency $f_{max}$ & 3 GHz\\
Bandwidth $B$ & 1.2 GHz\\
Sampling frequency $f_s$ & 1 MHz\\
Chirp duration $T_c$ & 0.6 ms\\
Num of samples per chirp $N_s$ & 600\\
Num of virtual receivers $N_r$ & 12\\
Frame rate $f_F$ & 12 FPS\\
\hline
\end{tabular}
\end{table}

\section{Methodology}
\label{Three}
\subsection{Radar System and Algorithms}

\subsubsection{Through-Wall Radar}
The through-wall radar used in this work is a portable FMCW radar developed by our team. The specific radar configuration is shown in Table \ref{tab:table1}.

The raw ADC data is a cube with dimensions ($N_s$, $N_r$, $N_c$). For a target at distance  $r$, the intermediate frequency reflects the magnitude of the distance value, which can be represented as:
\begin{equation}
\label{eq1a}
f_{IF} = \frac{2Sr}{c} ,
\end{equation}
where $S$ represents the slope of the chirp, and $c$ denotes the speed of light. The closest differentiable distance of two objects are defined as the range resolution of a radar, which can be derived from \eqref{eq1a} as follows:
\begin{equation}
\label{eq2a}
R_{res} = \frac{c}{2ST_c} = \frac{c}{2B}.
\end{equation}
Here, $T_c$ represents the duration of a chirp, and $B$ represents bandwidth. So, for our through-wall radar, its range resolution is 12.49 cm.

To estimate an object's angle, the radar needs at least two receiving antennas separated by a distance $d$. When an echo reflected from a target at an angle $\theta$ returns, it's captured by both receiving antennas. However, the signal reaching the second receiving antenna has to travel an additional distance of $d\sin \theta$, resulting in a phase difference $\omega$ between the two received signals. Therefore, the target angle can be determined by the following equation:
\begin{equation}
\label{eq3a}
\theta = \sin^{-1} \left ( \frac{\omega \lambda  }{2\pi d}  \right ),
\end{equation}
where $\lambda$ represents the wavelength. Hence, phase information is particularly crucial for radar target detection. We will further elaborate on this point in the subsequent sections to illustrate its significance.

\subsubsection{Traditional Signal Processing Algorithms}
\label{dft}
The Fourier Transform is a crucial concept in radar signal processing. This section extensively discusses the mathematical background and limitations of the Discrete Fourier Transform.

\begin{figure}[!t]
\centering
\subfloat[]{\includegraphics[width=1.72in]{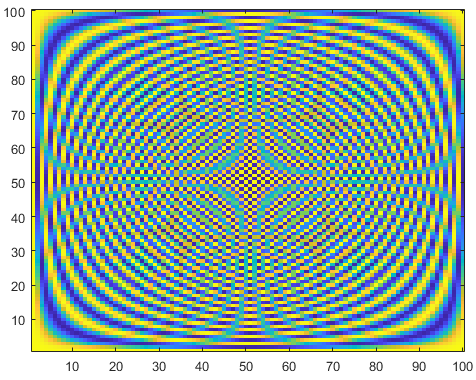}%
\label{fig_3a}}
\hfil
\subfloat[]{\includegraphics[width=1.72in]{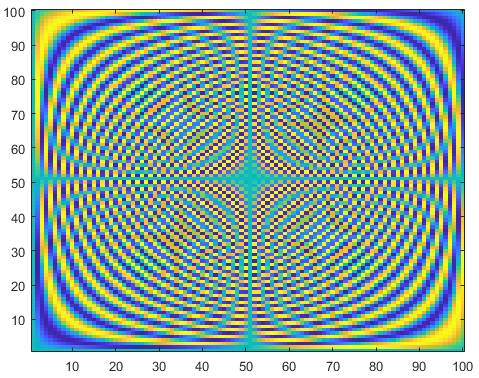}%
\label{fig_3b}}
\caption{The visualization result of the orthogonal basis parameters with a size of (N, N). (a) corresponds to the real part, while (b) corresponds to the imaginary part.}
\label{fig_3}
\end{figure}

For a sequence $\left \{ x\left [ n \right ]  \right \}_{0\le n< N}$, it forms an $N$-dimensional Euclidean space $\mathbb{C}^n$, the inner product of two signals is defined as:
\begin{equation}
    \label{eq4a}
    \left \langle a,b \right \rangle = \sum_{n=0}^{N-1} a_{i} \cdot b_{i} .
\end{equation}
Providing a set of orthogonal bases on $\mathbb{C}^n$ :
\begin{equation}
    \label{eq5a}
    \left \{ e_{k}\left[ n \right ] = e^{i\frac{2\pi }{N} kn}   \right \}_{0\le k< N} ,
\end{equation}
the decomposition of signal $x$ on this set of orthogonal bases yields:
\begin{equation}
    \label{eq6a}
    x = \sum_{k=0}^{N-1}\frac{\left \langle x,e_{k}  \right \rangle}{\left \| e_{k}  \right \|^{2}}e_{k} .
\end{equation}
Hence, let $\hat{x}$ be the Fourier transform of $x$: $\hat{x}$ = $\mathcal{F}x$, it can be expressed as follows:
\begin{equation}
    \label{eq7a}
    \hat{x} \left [ k \right ]=\left \langle x, e_{k}\right \rangle =\sum_{n=0}^{N-1}x[n]e^{-i\frac{2\pi }{N} kn} .
\end{equation}
Ultimately, \eqref{eq7a} can be understood as the discrete Fourier transform, where $\hat{x}$ being essentially the components of the discrete signal $x$ on the orthogonal bases $\left \{ e_{k}  \right \}$. Fig. \ref{fig_3}\subref{fig_3a}, \ref{fig_3}\subref{fig_3b} correspond to the real and imaginary parts of this set of bases, respectively. 

However, DFT as a fixed linear transformation shows significant room for improvement when dealing with through-wall data heavily affected by wall and obstacles, which can be summarized as follows: 1) \textbf{Data-driven transformation kernels.} Due to the lack of adaptive capability in the parameters of the Fourier transform, as depicted in Fig. \ref{fig_3}, its optimality is constrained by the use of this type of spectrum graph. By integrating the signal processing pipeline into the learning optimization process of the network, it becomes possible to adaptively reconstruct the uniformly discretely distributed features across the entire observable frequency domain. 2) \textbf{Finer features.} According to \eqref{eq2a}, radar resolution is solely determined by the bandwidth. When two targets are very close together, relying on the coarse features provided by DFT makes it difficult to distinguish between the two. Networks, however, can complement more nuanced texture features beyond DFT's limitation into the output heatmap. This significantly increases the likelihood of the subsequent network detecting both targets.

\subsection{Network Details}
\begin{figure*}[!t]
\centering
\includegraphics[width=7.1in]{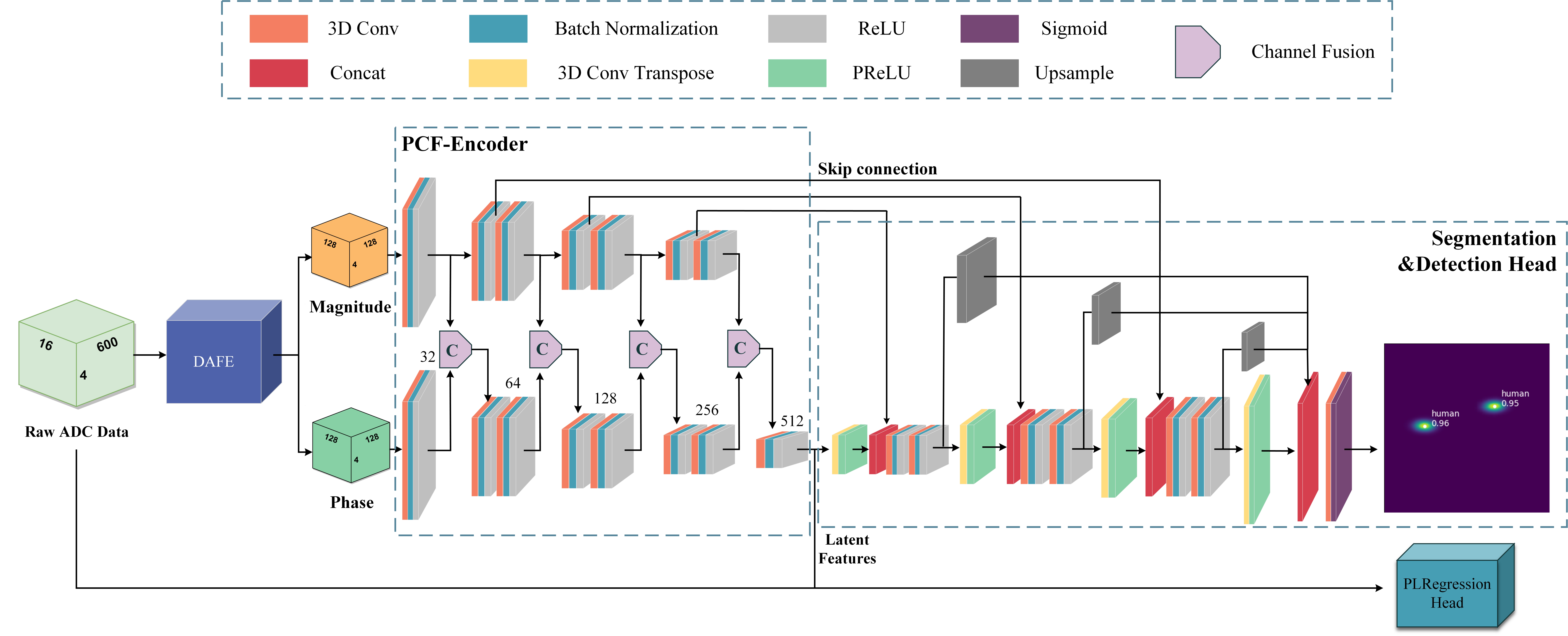}
\caption{The overall architecture of the TWP-CNN.}
\label{fig_4}
\end{figure*}

The network we proposed will be detailed in this section. As shown in Fig. \ref{fig_4}, our network uses the raw ADC cube from the radar as input data, without any preprocessing. Instead, we employ DFT-based adaptive feature extraction (DAFE) convolutional layers for signal processing. It will learn to transform ADC data into a specific domain, including the frequency domain, while preserving both the magnitude and phase information of the data. Subsequently, through a dual-stream encoder, the phase information is embedded into the feature maps extracted by the network via channel fusion. Then, the decoder reconstructs the confidence heatmap of the space behind the wall from the extracted high-level semantic information, highlighting the targets. Finally, leveraging the location-based non-maximum suppression (L-NMS) and object location similarity (OLS) proposed in \cite{wang2021rodnet}, post-processing is applied to the confidence heatmap to obtain the target detection and evaluation results. Additionally, apart from segmentation and detection, we introduced a coordinate regression (PCR) module. This module enhances the model's ability to generalize, facilitating a better understanding of target features. It utilizes the phase information of the original data, and the network's latent representation to output the distance and angle coordinates of the targets. Compared to utilizing amplitude data only, higher accuracy can be attained.

\subsubsection{DAFE}
\label{dafe}
\begin{figure}[!t]
\centering
\includegraphics[width=3.4in]{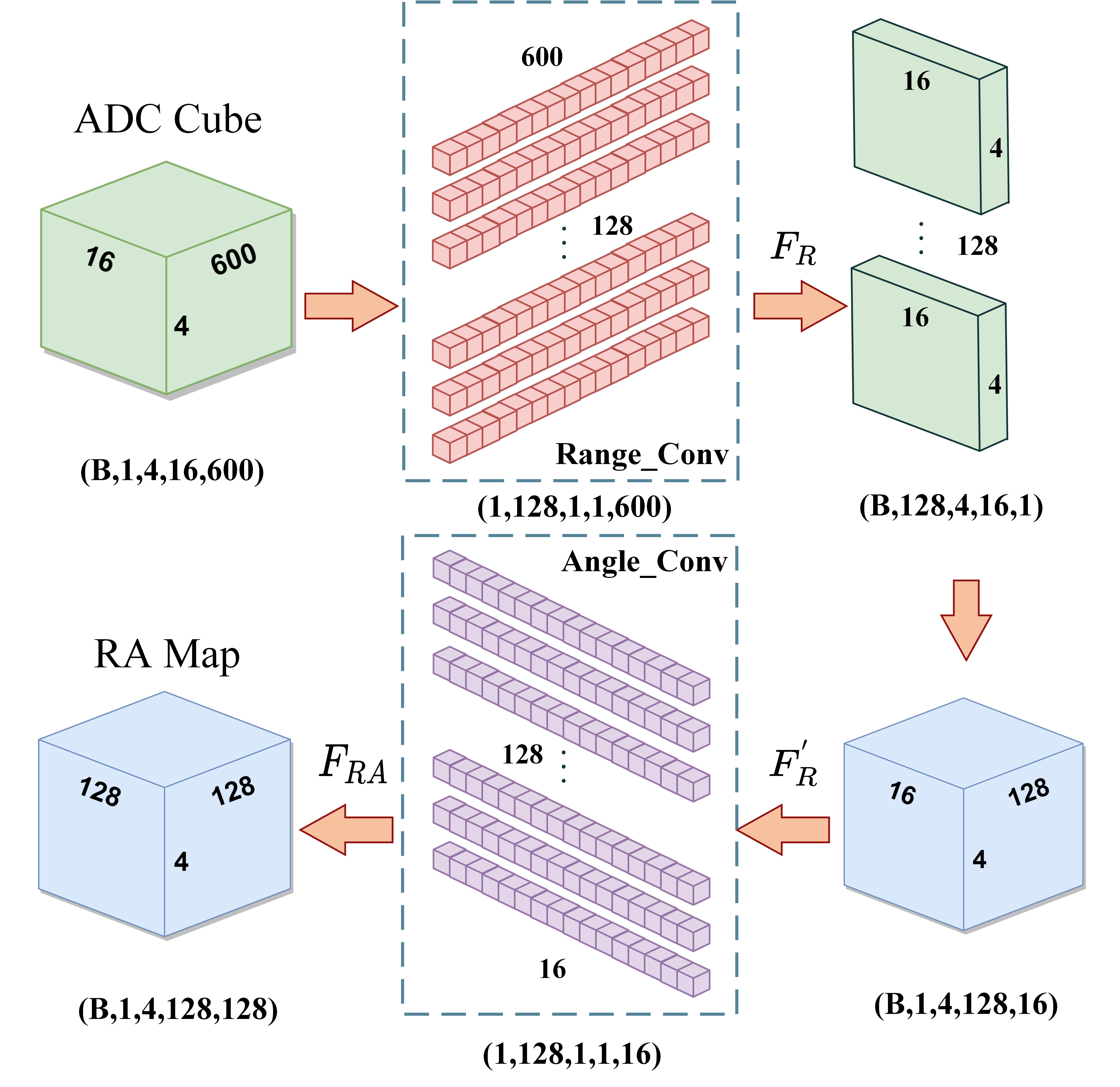}
\caption{DFT-based adaptive feature extraction module (DAFE).}
\label{fig_5}
\end{figure}

This module integrates the signal processing flow into the network architecture, eliminating the need for explicit features in the network input or preprocessed data. Instead, the network learns implicit features that can approximate the optimal representation. Hence, it is possible to use 3D convolutions instead of specific DFTs and initialize the weights of the 3D convolution with the parameters of the Fourier transform basis. As depicted in Fig. \ref{fig_5}, the size of the raw ADC data $D_{raw}$ is (B, 1, 4, 16, 600), where B represents the batch size, 4 signifies the number of chirps per frame, 16 denotes the number of virtual receive channels in the radar, and 600 represents the number of intermediate frequency signal sampling points. The first step involves conducting range convolution $H_{r}(\cdot )$ along the sampling point dimension, with a kernel size of (1, 128, 1, 1, 600):
\begin{equation}
    \label{eq8a}
    F_{R} = H_{r}(D_{raw}). 
\end{equation}
The size of $F_{R}$ is (B, 128, 4, 16, 1), where the data along the second dimension represents the distance spectrum information.

Usually, initializing the convolution kernel weights using parameters from the Discrete Fourier Transform (DFT) kernel is reasonable. However, we opted for the Chirp-Z Transform (CZT) parameters due to the maximum detection range of 75 meters in our low-frequency through-wall radar system, where most through-wall scenarios have longitudinal distances within 15 meters. Therefore, utilizing DFT to extract frequency spectra along the distance axis would introduce significant amounts of irrelevant spectral information, causing interference and severely compressing the spatial bandwidth of useful frequencies. Additionally, our radar system employs a coaxial line length of 1 meter with a round-trip transmission distance of 2 meters. Hence, the section corresponding to the first two meters on the distance axis also lacks information. Adopting CZT effectively resolves these issues.

\begin{figure}[!t]
\centering
\subfloat[]{\includegraphics[width=3.4in]{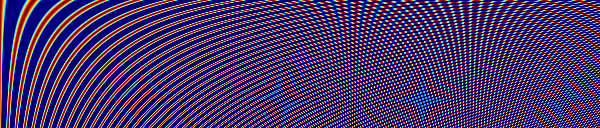}%
\label{fig_6a}}
\hfil
\subfloat[]{\includegraphics[width=3.4in]{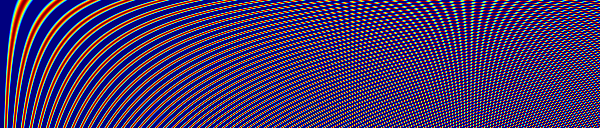}%
\label{fig_6b}}
\hfil
\subfloat[]{\includegraphics[width=1.7in]{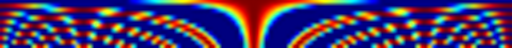}%
\label{fig_6c}}
\hfil
\subfloat[]{\includegraphics[width=1.7in]{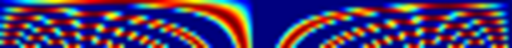}%
\label{fig_6d}}
\caption{Visualization diagram of the convolutional kernel parameters in the DAFE module. (a) represents the real part of the weights for distance convolution, while (b) corresponds to the imaginary part, with dimensions of (128, 600). (c) represents the real part of the weight for angular convolution, while (d) corresponds to the imaginary part, with dimensions of (16, 128).}
\label{fig_6}
\end{figure}

CZT, a generalization of DFT, differs from DFT by uniformly sampling multiple divisions on the unit circle in the Z-plane. CZT enables sampling along a segment of a spiral on the Z-plane for equiangular sampling.The sampling points can be expressed as:
\begin{equation}
    \label{eq9a}
    z_{k} = AW^{-k}, k=0,1,...,N-1 ,
\end{equation}
where N represents the total number of sampling points, A stands for the starting point position: $A = A_0e^{j\theta _0}$, and W denotes the ratio between points: $W = W_0e^{j\phi  _0}$. In this approach, the CZT parameters are set for frequency domain transformation of the section ranging from 2.2 to 10 meters, thus $A_0$ and $W_0$ are both set to 1. The final parameter expression for the Chirp-Z Transform is:
\begin{equation}
    \label{eq10a}
    z_k^{-n} = e^{-j2\pi\frac{nf_{start}}{f_{s}}}e^{-j2\pi \frac{nk(f_{end} - f_{start})}{Nf_s} } ,
\end{equation}
where $f_s$ represents the sampling rate, $f_{start}$ corresponds to the frequency value at 2.2 meters, and $f_{end}$ corresponds to the frequency value at 10 meters. These values can be adjusted according to different application scenarios, providing high flexibility. The visualization of these parameters is depicted in Fig. \ref{fig_6}, where Fig. \ref{fig_6}\subref{fig_6a} represents the real part of this complex weight matrix, and Fig. \ref{fig_6}\subref{fig_6b} represents the imaginary part. It's evident that it differs significantly from the parameter matrix of DFT.

Then, after reshaping $F_{R}$ into $F_{R}^{' }$, angular convolution is applied to $F_{R}^{' }$. 
\begin{equation}
    \label{eq11a}
    F_{RA} = H_{a}(F_{R}^{' }),
\end{equation}
where $H_{a}(\cdot )$ denotes angular convolution operation. The convolution layer is initialized using standard DFT parameters. The weights require a swap between the first half and the second half to shift the zero frequency component to the center, akin to an fftshift operation. However, in our case, we shift the parameters of the transformation base, whereas fftshift involves shifting the resulting data. Essentially, there's no difference in essence, just in how it's implemented in code. Fig. \ref{fig_6}\subref{fig_6c}, \ref{fig_6}\subref{fig_6d} display the real and imaginary parts of the parameter matrix for the angular convolution. The resulting (128, 128) image obtained after distance and angular convolutions is referred to as the RA Map. At this point, the RA Map is a complex tensor. We simultaneously extract its magnitude and phase values, utilizing them as the output data for DAFE. However, at this stage, the RA Map remains coarse and requires further processing through subsequent networks to eliminate noise and extract target features.
\subsubsection{PCF-Encoder} 
In radar target detection processes, typically, only the amplitude values are used as input data, while the phase information is often discarded. In situations with good signal-to-noise ratios, this doesn't significantly affect performance as the amplitude spectrum alone can identify targets. Nevertheless, in complex environments with high noise levels, distinguishing targets from the amplitude spectrum alone becomes challenging. Yet, utilizing phase information in such scenarios can mitigate the degradation and improve performance to a certain extent. In the meantime, objects at varying distances and angles induce differences in phase, aiding in a more precise localization of their distance and angle. 

\begin{figure}[!t]
\centering
\includegraphics[width=3.4in]{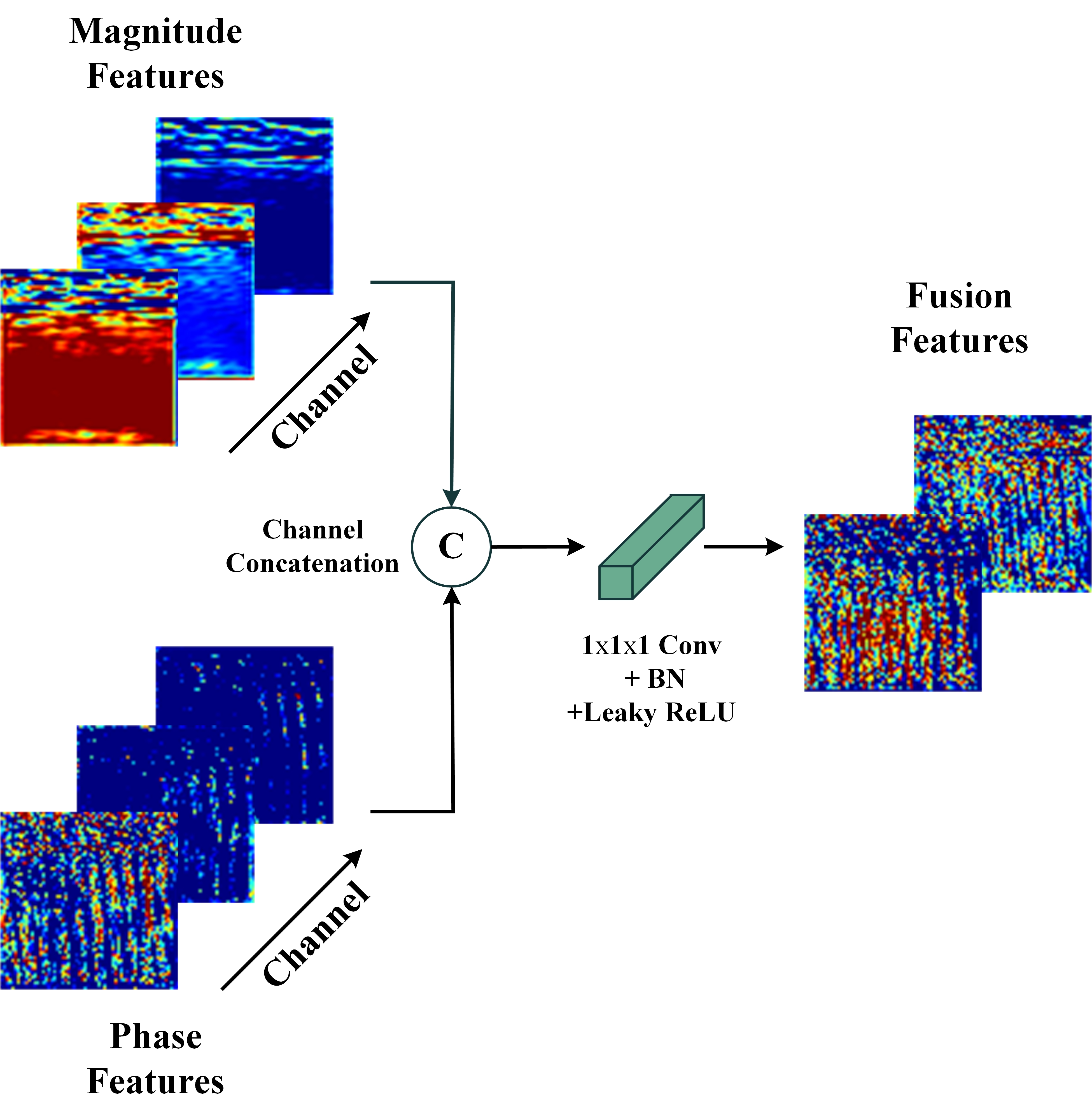}
\caption{Channel Fusion Module.}
\label{fig_7}
\end{figure}

Therefore, upon obtaining the RA Map output from DAFE, we perform fusion and feature extraction on its amplitude and phase data using the phase-based channel fusion encoder (PCF-Encoder) proposed in our method. However, phase maps are inherently difficult to discern and lack intuitiveness. Directly adding them to amplitude maps or fusing them spatially is unreasonable. In this work, as shown in Fig. \ref{fig_4}, we extract features separately for amplitude and phase information through two branches. Then, we perform channel-wise concatenation on the multi-scale features extracted from the two branches, merging their high-dimensional representations along the channels. The channel fusion operation is illustrated in Fig. \ref{fig_7}, where the concatenated feature maps undergo a simple 1x1x1 convolutions, batch normalization, and Leaky ReLU activation to achieve channel compression and information fusion. This process effectively integrates phase information into the high-dimensional representation of the network.

\subsubsection{Phase-based Coordinate Regression Head} 
\begin{figure}[!t]
\centering
\includegraphics[width=3.4in]{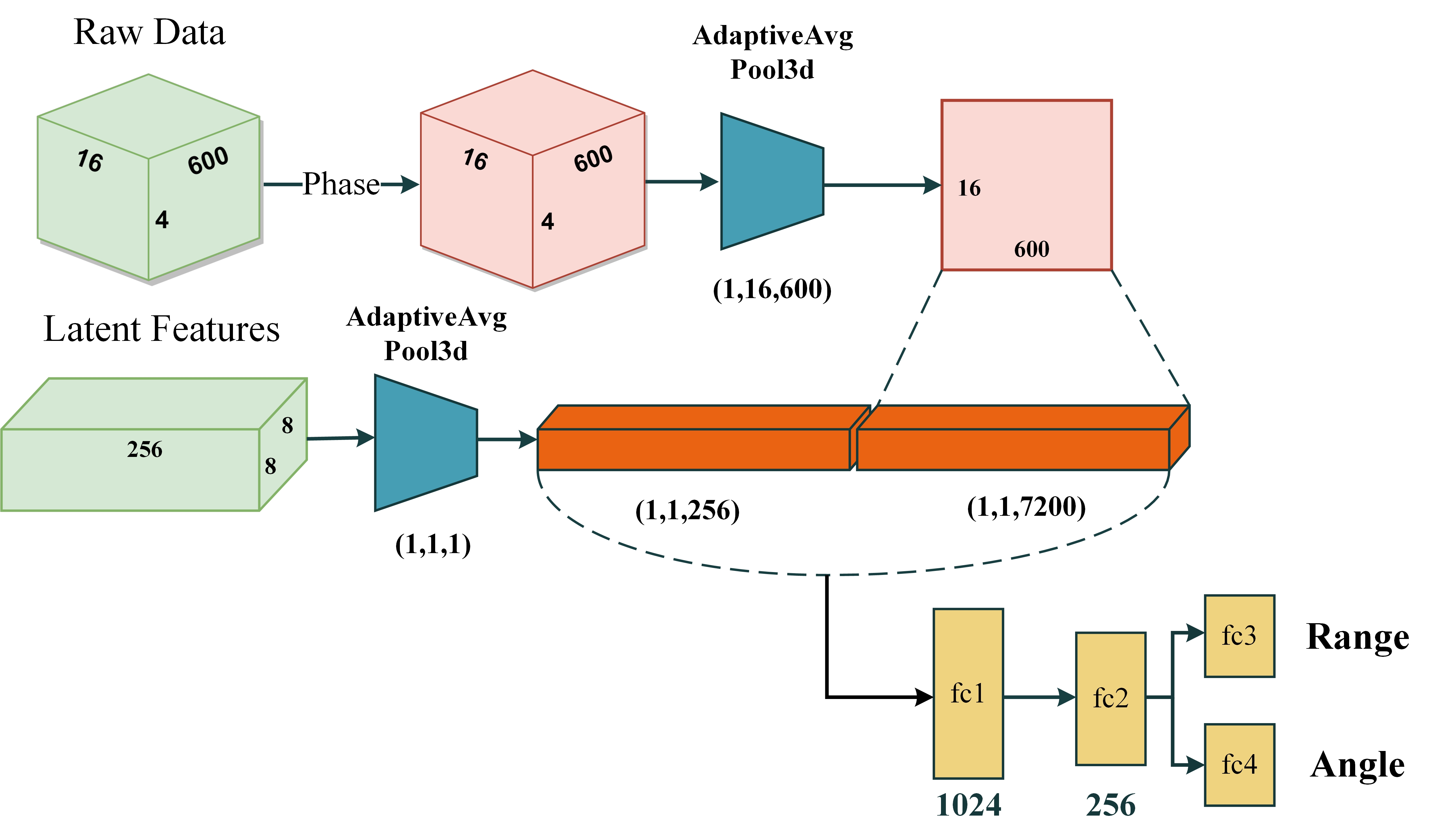}
\caption{Phase-based Coordinate Regression Module.}
\label{fig_8}
\end{figure}

In this network, a module called the Phase-based Coordinate Regression (PCR) Module is proposed to obtain the distance and angle values corresponding to the targets. Fig. \ref{fig_8} depicts the raw ADC data and latent features extracted by the Encoder being used as input data for this module. Transforming the raw ADC data into phase breaks through the accuracy bottleneck, due to the sensitivity of FMCW radar's phase to small object movements. Then, to mitigate the impact of variations in target positions across multiple frames, an adaptive average pooling is applied to merge and reduce the dimensions of the data from four frames to a (1, 16, 600) data. After that, it's flattened into a (1, 1, 7200) vector. The other input for this module is the latent feature from the end of the encoder, sized (8, 8, 256). Similarly, an adaptive average pooling is employed to obtain a (1, 1, 256) vector from this data. The two vectors are concatenated and then fed together into two fully connected layers(fc1, fc2). Subsequently, separate predictions for the target's distance and angle values are made using two distinct fully connected layers (fc3, fc4).

\subsubsection{Loss Function}
Our task is divided into two components: segmentation and regression. The segmentation task involves binary heatmap prediction for Range-Angle, while the regression task predicts the distance and angle values for the targets. In the segmentation process, we utilized Focal Loss and Charbonnier Loss. While cross-entropy loss functions are commonly employed in network training, they tend to face challenges when dealing with significantly imbalanced positive and negative sample quantities. In our specific task, the number of positive samples (human bodies) is notably lower than that of negative samples (background). Furthermore, due to substantial noise during penetration, targets often exhibit faint and indistinguishable characteristics. Hence, employing Focal Loss better suits our task requirements and demonstrates superior performance. 

The Focal Loss is based on the cross-entropy loss function, and its formula is as follows:
\begin{equation}
    \label{eq12a}
    FL(p_t) = -\alpha_t \cdot (1 - p_t)^\gamma \cdot \log(p_t),
\end{equation}
where $p_t$ denotes the probability predicted by the model, $\alpha_t$ is utilized to adjust the ratio between the losses of positive and negative samples within the range of 0 and 1, and $(1 - p_t)^\gamma$ is employed to diminish the loss contribution from easily classifiable samples while amplifying the portion of losses from challenging-to-classify samples. We obtained the best performance by setting $\alpha_t$ to 0.8 and $\gamma$ to 2. Furthermore, drawing inspiration from related work in the domain of super-resolution and image generation, we noticed the presence of low-confidence artifacts on the binary heatmaps obtained. As a result, we augmented the segmentation loss with a Charbonnier Loss to enhance the quality of the heatmaps. The formula is as follows:
\begin{equation}
    \label{eq13a}
    CL(y- \hat{y}) = \sqrt{(y- \hat{y})^{2} + \epsilon ^{2}} ,
\end{equation}
where $y$ represents the network's output heatmap and $\hat{y}$ represents the corresponding label. For the regression part, we utilize L1 loss separately to compute the loss for distance and angle values. The loss function for our entire task is formulated as follows:
\begin{equation}
    \label{eq14a}
    L_{total} = FL(y,\hat{y} ) + CL(y,\hat{y} ) + \alpha L1(r,\hat{r}) + \beta L1(a, \hat{a} ) ,
\end{equation}
where $\alpha$ and $\beta$ represent the weights for the regression loss, both set to 0.1. The weight for the segmentation part is 1. Additionally, $r$ and $\hat{r}$ denote the predicted and true values for distance, while $a$ and $\hat{a}$ represent the predicted and true values for angle.

\section{Experiments}
\label{Four}
\subsection{Realistic Through-Wall (RTW) Datasets}
\label{dataset}
FMCW radar datasets are widely present in the field of autonomous driving \cite{caesar2020nuscenes, ouaknine2021carrada, mostajabi2020high, wang2021rethinking, rebut2022raw}. In comparison to cameras and LiDAR, radar data in these datasets typically operates at relatively lower frequencies. Additionally, radar data is often represented in forms such as range-azimuth-Doppler tensor, range-azimuth view, range-Doppler view, or point cloud, with minimal utilization of raw ADC data.

Given our network's reliance on raw ADC data from the radar, existing through-wall radar datasets lack this crucial information, relying instead on preprocessed radar data \cite{zhengliang2021dataset}, \cite{tian2022uwb}. Hence, utilizing our proprietary through-wall radar system, we've assembled a real and challenging through-wall radar dataset, termed RTW (Realistic Through-Wall) Dataset. This dataset serves to evaluate the capabilities of our methods in practical applications. The collected data does not involve self-constructed brick walls. Instead, it focuses on common concrete walls found in actual buildings. These walls commonly contain metallic mediums like wires and pipelines distributed non-uniformly, resulting in a relatively low signal-to-noise ratio in the data. 

\begin{figure}[!t]
\centering
\includegraphics[width=3.4in]{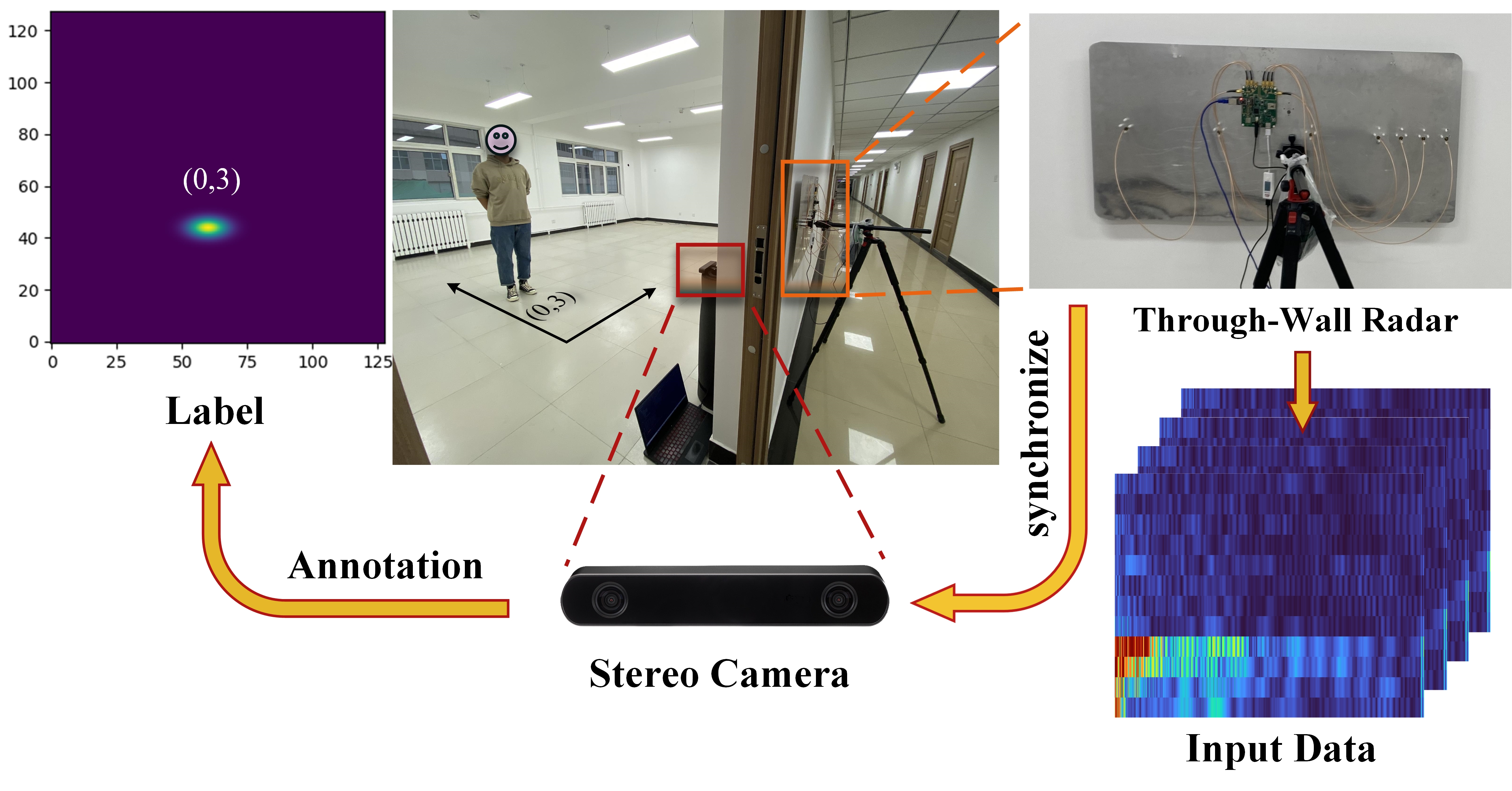}
\caption{Illustration for RTW Dataset collection. A target is positioned at coordinates (0, 3). Data is simultaneously collected by the camera and through-wall radar, obtaining raw ADC data and label values.}
\label{fig_9}
\end{figure}

\begin{figure}[!t]
\centering
\includegraphics[width=3.4in]{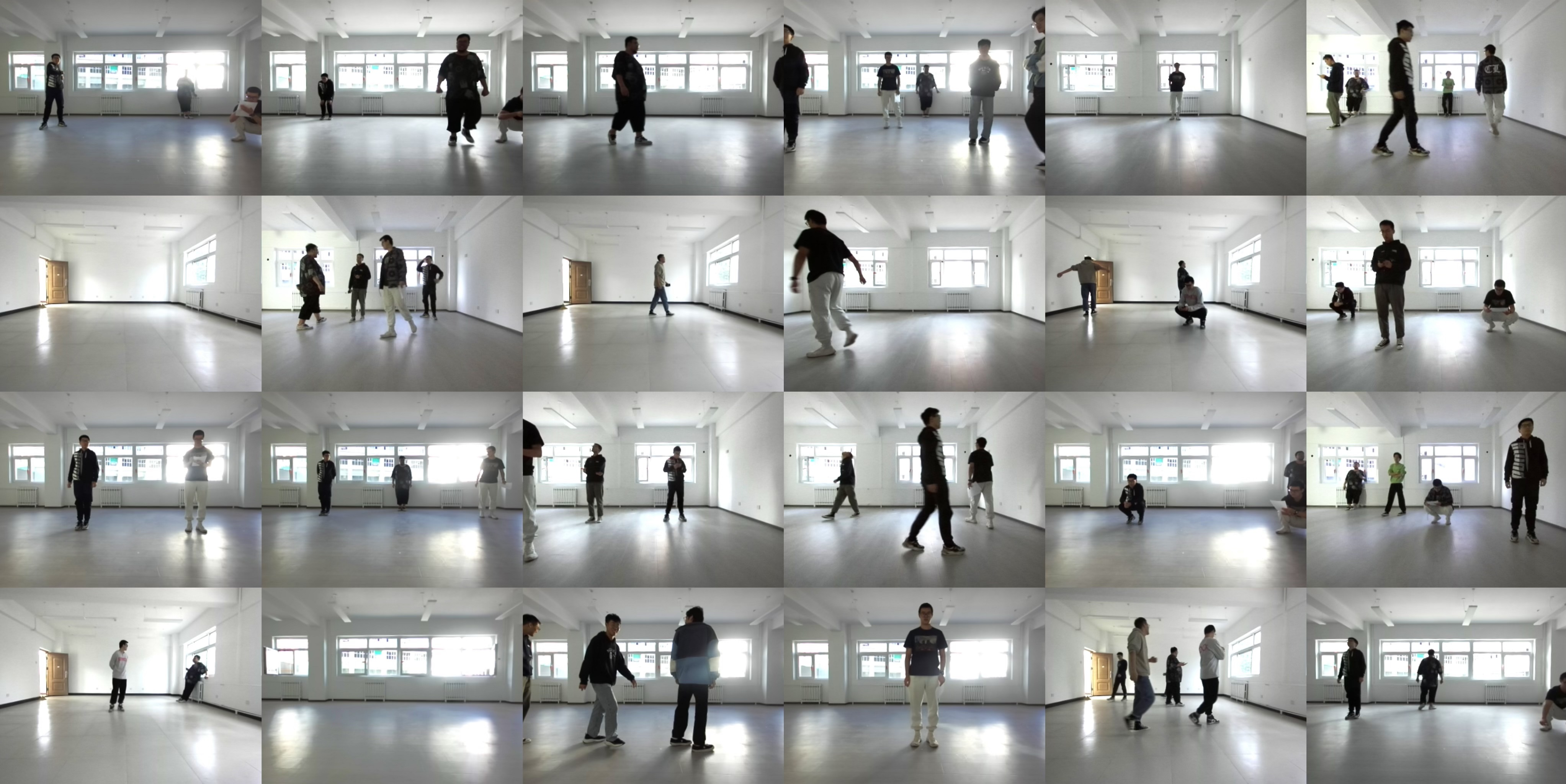}
\caption{Some samples from the dataset.}
\label{fig_10}
\end{figure}

\begin{figure}[!t]
\centering
\includegraphics[width=3.4in]{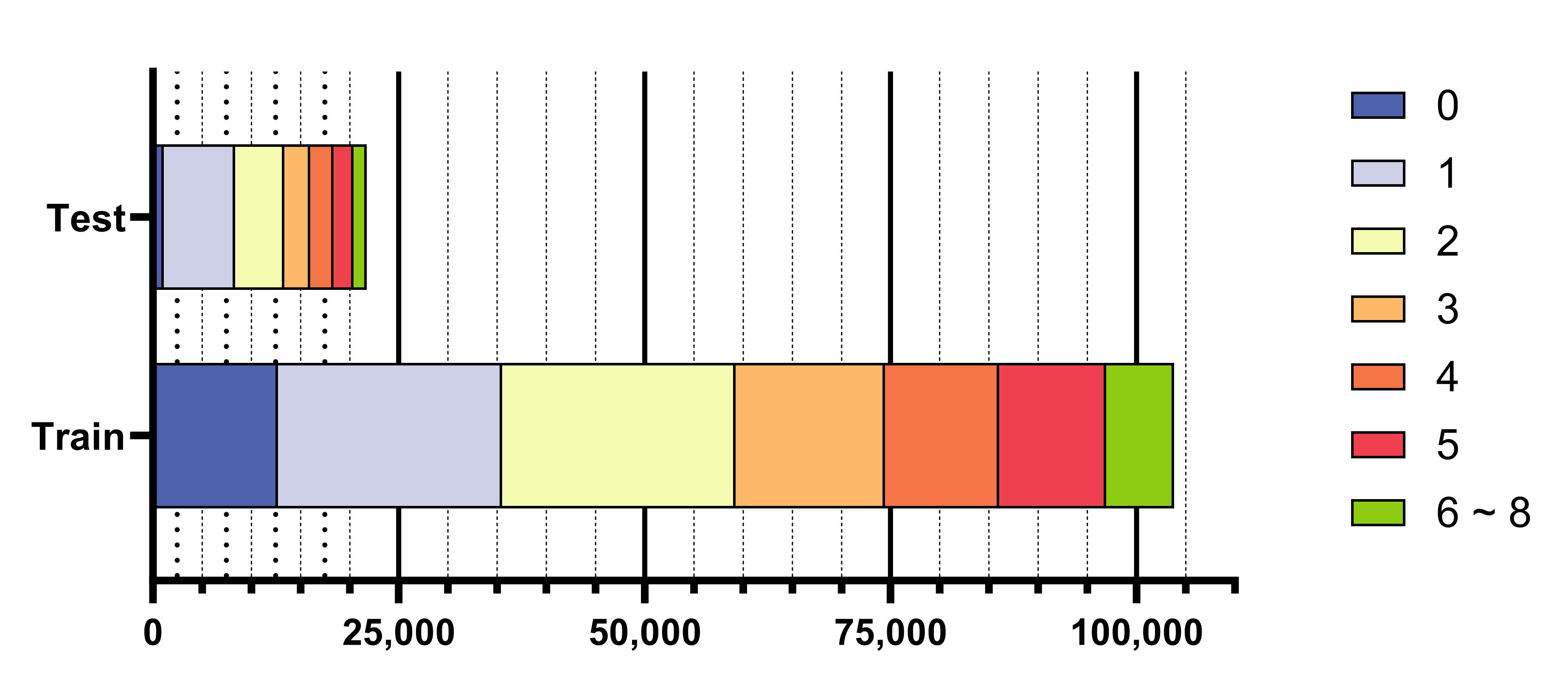}
\caption{The distribution of the dataset. Each bar represents the number of frames containing a specific quantity of targets.}
\label{fig_11}
\end{figure}

\begin{table}[!t]
\caption{Statistics of The Dataset\label{tab:table2}}
\centering
\begin{tabular}{c   c   c}
\hline
Dataset & Train & Test\\
\hline
Number of frames  & 45K & 8600\\
Number of sequences  & 32 & 6\\
Duration & 1h53m & 19m48s\\
Number of annotations  & 104k & 22k\\
Number of simple sequences  &  8  & 2\\
Number of medium sequences  &  15  &  2\\
Number of hard sequences  &  9  &  2\\
\hline
\end{tabular}
\end{table}

Our data collection setup comprises a through-wall radar, a stereo camera, and a PC, as illustrated in Fig. \ref{fig_9}. The stereo camera is positioned indoors, affixed to the wall at a height of 1 meter above the ground, capturing continuous indoor images at a rate of 30 frames per second (FPS). Subsequently, utilizing object detection networks and depth estimation, obtaining the two-dimensional coordinates of each target within frames to generate labels for the dataset. Inspired by \cite{wang2021rodnet},\cite{cao2017realtime}, both methods utilize heatmap representation to indicate the targets' positions. In the RA Map, each pixel's value signifies the probability of a target's presence at that specific location, employing a Gaussian distribution to determine values around the target point. We followed this approach to generate annotated heatmap labels.

Furthermore, the through-wall radar is positioned outdoors, aligned with the camera, situated at a height of 1.5 meters above the ground and placed flush against the wall. The radar and camera initiate data collection simultaneously under PC control. Each frame of radar data is time-stamped to match the closest temporal label data. The obtained ADC data serves as the input data for the network, devoid of any preprocessing.

The dataset comprises data collected from multiple scenes across three different rooms, totaling approximately 2 hours in duration. Fig. \ref{fig_10} displays some examples from the dataset. The collection of the training and testing sets occurred independently, devoid of temporal correlations or identical scenes, thereby preventing data leakage. The statistical data of the dataset is presented in Table \ref{tab:table2}. The training set comprises 85\% of the dataset, with 20\% of the training set further partitioned for validation set. Subsequently, the testing set was divided into three different difficulty levels based on the number of targets, signal-to-noise ratio, and wall thickness. This categorization allows for a more detailed comparison of experimental results. Fig. \ref{fig_11} displays the number of frames containing a specific quantity of targets in both the training and testing sets. 

\begin{table*}[!t]
\caption{Through-Wall Human Body Detection Results\label{tab:table3}}
\centering
\begin{tabular}{c |c c | c c |c c | c c |c c | c c}

\hline
\multirow{2}{*}{Models} &
\multicolumn{4}{c|}{Simple} &
\multicolumn{4}{c|}{Medium} &
\multicolumn{4}{c}{Hard}\\
\cline{2-13}
     & AP & AP$_{50}$ & AR & AR$_{50}$ & AP & AP$_{50}$ & AR & AR$_{50}$ & AP & AP$_{50}$ & AR & AR$_{50}$\\

\midrule
FCN\cite{li2020human}  & 63.18 & 67.52 & 67.43 & 70.29 & 42.72 & 52.12 & 50.69 & 56.88 & 28.43 & 35.07 & 32.94 & 37.17\\
Improved U-Net\cite{pan2023multi}  & 68.59 & 73.46 & 70.92 & 74.31 & 44.51 & 55.87 & 54.21 & 59.32 & 29.11 & 37.50 & 34.67 & 38.42\\

FCN+DAFE  & 75.64 & 79.65 & 81.95 & 84.52 & 49.85 & 61.89 & 58.09 & 65.42 & 36.53 & 44.30 & 42.42 & 47.69\\
Improved U-Net+DAFE  & 78.16 & 84.24 & 83.67 & 87.38 & 52.49 & 64.37 & 61.89 & 67.58 & 37.05 & 45.62 & 43.74 & 48.28\\

\bottomrule
TWP-CNN(ours) & \textbf{88.24} & \textbf{95.41} & \textbf{91.90} & \textbf{97.25} & \textbf{64.18} & \textbf{78.07} & \textbf{71.92} & \textbf{80.00} & \textbf{46.59} & \textbf{52.96} & \textbf{49.81} & \textbf{54.51}\\
\hline
\end{tabular}
\end{table*}

\begin{figure*}[!t]
\centering
\includegraphics[width=7.1in]{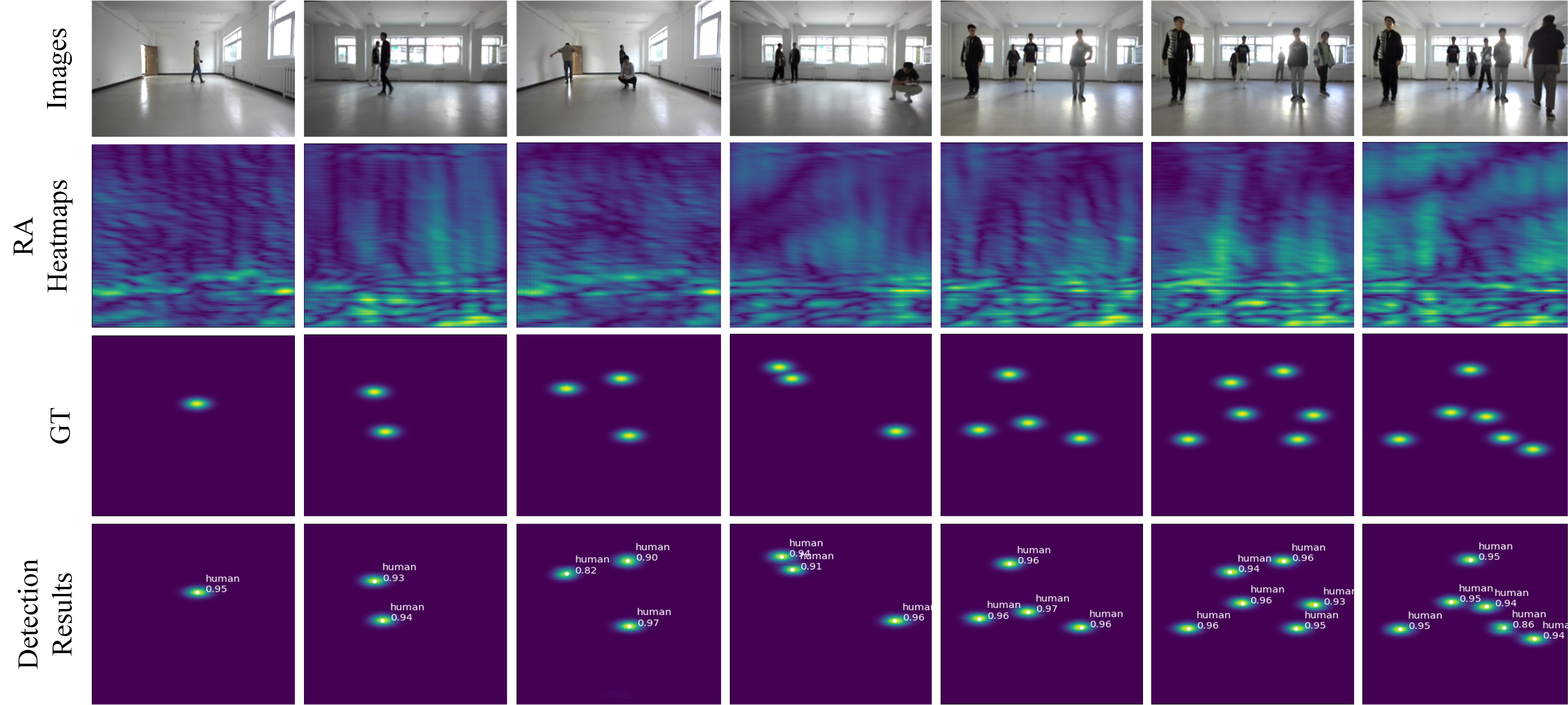}
\caption{ Some results from the test sets after passing through the network, encompassing stationary and moving targets. The first row depicts images captured by the camera, the second row exhibits the RA heatmap derived from the original ADC data after DAFE processing. The third row represents the GT (Ground Truth) heatmap generated from labeled values. Lastly, the fourth row illustrates the segmented RA heatmap and detection results produced by the network. This demonstrates our network's exceptional through-wall detection performance across multiple targets and scenes.}
\label{fig_12}
\end{figure*}

\subsection{Training Details and Evaluation Metrics.}
We compare our TWP-CNN with models from two other papers \cite{li2020human, pan2023multi} focusing on the same task. Considering that the SNR level of the dataset used for evaluation has surpassed the processing capabilities of traditional detection algorithms, we did not compare our approach with traditional algorithms in this experiment. When training TWP-CNN, we employ a cosine annealing learning rate ranging from $1\times 10^{-4} $ to $1\times 10^{-5} $, with a $T_0$ of 10 and a $T_{mul}$ of 3. The total training span 130 epochs, utilizing the Adam optimizer. The entire training and testing procedures are conducted on an RTX 3090 GPU. The comparative models are replicated based on the structures and training methodologies described in their respective papers.

According to RODNet's evaluation criteria \cite{wang2021rodnet}, we use OLS (object location similarity) instead of IoU (Intersection over Union) to determine the similarity between the output results and the Ground Truth. Subsequently, we compute the Average Precision (AP) and Average Recall (AR) at various thresholds ranging from 0.5 to 0.9. Different thresholds represent detection results within varying error tolerances.

\subsection{The Through-Wall Human Body Detection Results.}
We train our network and the two comparative networks, then we evaluate them using different difficulty levels of the test sets. We quantitatively compare them using the AP and AR metrics. The results are showcased in Table \ref{tab:table3}. Our TWP-CNN achieves 88.24\% AP and 91.90\% AR on simpler data (e.g., with only a single target), surpassing the performance of the other two methods by a significant margin. Compared to FCN on data of intermediate difficulty, our approach shows a 52.23\% improvement in AP and a 41.88\% improvement in AR. In comparison to U-Net, there's an 44.19\% increase in AP and a 32.67\% increase in AR. When tested with hard data, both FCN and U-Net exhibit a significant decrease in AP and AR metrics. Nevertheless, we still manage to achieve approximately 50\% AP and AR.

In Fig. \ref{fig_12}, we showcase various test results, encompassing different scenarios, diverse quantities of individuals, and both static and dynamic targets. This effectively demonstrates the efficacy of our approach in real through-wall scenarios. The second row in the figure doesn't depict the network's input data or the RA Map after traditional radar preprocessing like DFT. Instead, it represents the RA heatmap output by the learned DAFE layer. While it may appear similar in form to the output of DFT processing, it is not the result of a Fourier transform. Instead, it represents the network-learned, finer, and more comprehensive feature representation. The results in the fourth column of the Fig. \ref{fig_12} indicate that our network is capable of processing and extracting finer features, enabling differentiation even when two individuals are in close proximity. The resolution is significantly enhanced. The results depicted in the figure also demonstrate that utilizing multi-frame texture features effectively detects stationary targets.

\begin{table*}[!t]
\caption{Ablation Studies of Network Performance by Several Modules\label{tab:table4}}
\centering
\begin{tabular}{c |c c c | c |c c c | c |c c c }

\hline
Models & DAFE & PCF-Encoder & PCR & AP & AP$_{50}$ & AP$_{75}$ & AP$_{90}$ & AR & AR$_{50}$ & AR$_{75}$ & AR$_{90}$\\

\midrule
\multirow{6}{*}{TWP-CNN} & 
 &  &  & 74.79 & 81.65 & 76.09 & 60.89 & 77.67 & 83.63 & 78.76 & 68.15\\
&$\surd$ &  &  & 86.05 & 88.26 & 87.14 & 76.13 & 89.53 & 93.74 & 90.61 & 78.25\\
&& $\surd$ &  & 83.87 & 87.80 & 85.47 & 75.29 & 85.99 & 89.39 & 87.02 & 79.15\\
&&  & $\surd$ & 85.37 & \underline{90.53} & 87.07 & 73.36 & 89.22 & 93.64 & 90.76 & 79.65\\

&$\surd$ & $\surd$ &  & \underline{86.48} & 88.60 & \underline{87.88} & \underline{77.30} & \underline{90.84} & \underline{95.88} & \underline{92.75} & \underline{80.38}\\
&$\surd$ & $\surd$ & $\surd$ & \textbf{88.24} & \textbf{95.41} & \textbf{90.18} & \textbf{73.61} & \textbf{91.90} & \textbf{97.25} & \textbf{93.13} & \textbf{81.27}\\
\hline
\end{tabular}
\end{table*}

\begin{figure*}[!t]
\centering
\subfloat[]{\includegraphics[width=1.72in]{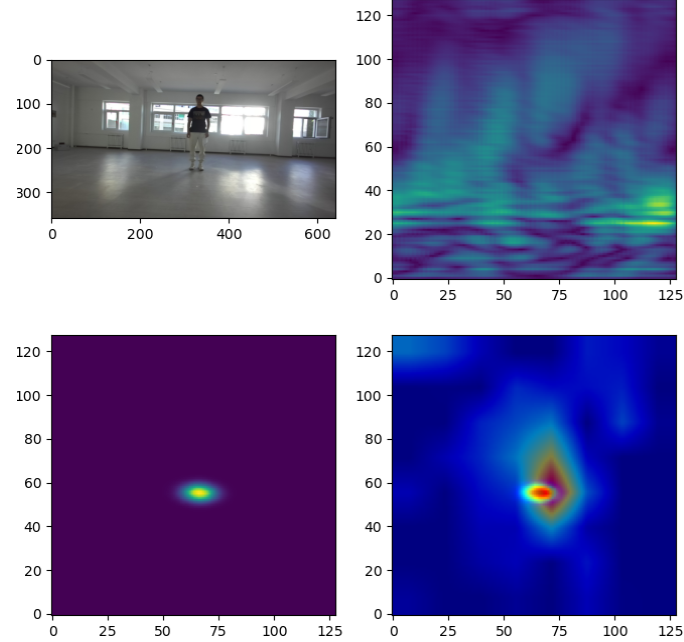}%
\label{fig_13a}}
\subfloat[]{\includegraphics[width=1.72in]{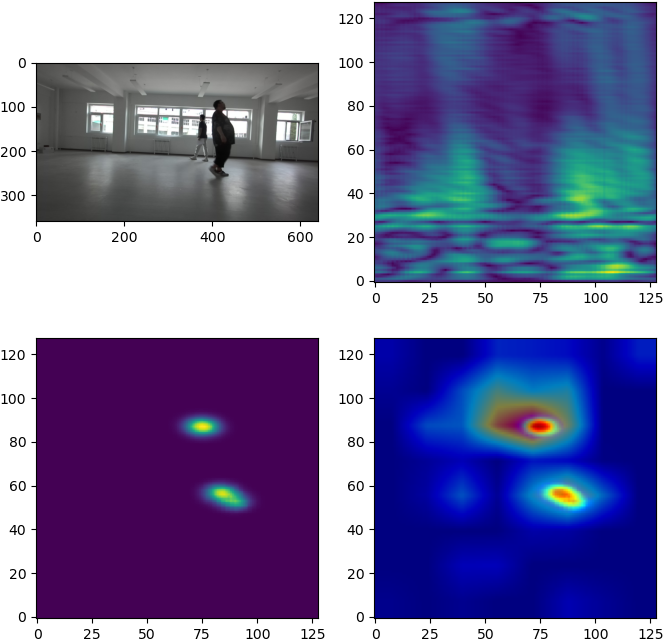}%
\label{fig_13b}}
\subfloat[]{\includegraphics[width=1.72in]{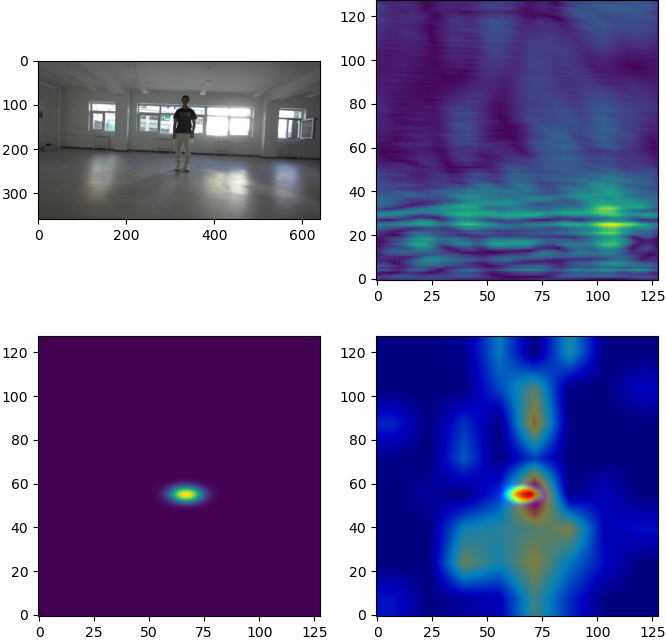}%
\label{fig_13c}}
\subfloat[]{\includegraphics[width=1.72in]{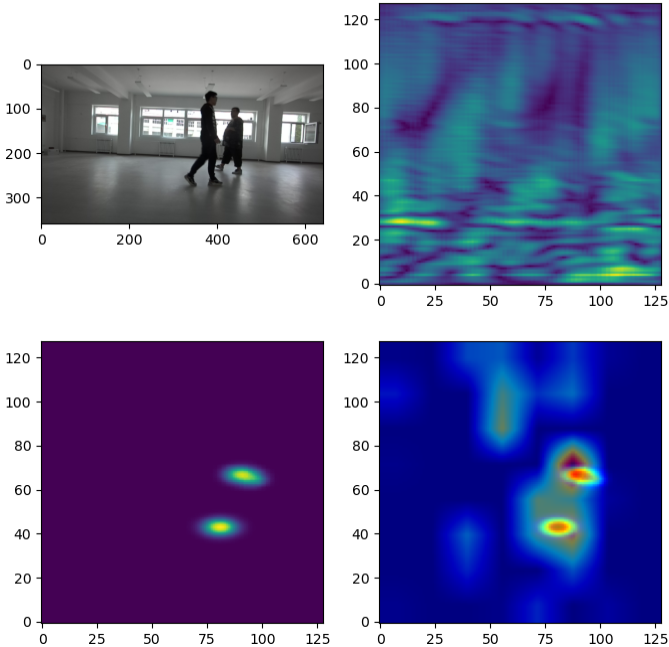}%
\label{fig_13d}}
\caption{GradCAM++ visualization results. In each image, the top-left corner displays the camera photo, the top-right corner showcases the RA Map output from DAFE, the bottom-left corner exhibits the network's segmentation prediction result, and the bottom-right corner presents the overlay of class activation maps  (CAM) on the segmentation result.}
\label{fig_13}
\end{figure*}

 \subsection{Ablation Experiments}
 \label{ablation}
In this section, we validate the effectiveness of the proposed modules and quantitatively analyze their contribution to the entire model.

DAFE is one of the most critical modules proposed by us. As discussed in \ref{dafe}, it directly handles raw ADC data, replacing the conventional fixed DFT with 3D complex convolutions. This enables the model to adaptively extract features from ADC by learning and optimizing the transformation matrix parameters. It significantly enhances the upper limit of the model when dealing with complex and challenging data. Moreover, it can be seamlessly embedded into other networks, significantly enhancing their performance. As shown in Table \ref{tab:table3}, after incorporating the DAFE module into two comparative experimental networks, FCN and Improved U-Net, both AP and AR metrics exhibited substantial improvements compared to the original networks. On the Hard test set, there was an increase of approximately 28\%, and a 15\% to 20\% improvement was observed on relatively simpler data. Furthermore, the input data of the original comparative network without the incorporation of the DAFE module is obtained through conventional preprocessing methods. The improvement in metrics resulting from the integration of DAFE once again demonstrates the superiority of our proposed learning-based signal processing approach over traditional preprocessing pipelines. In Table \ref{tab:table4}, comparing the base models without proposed modules, we observed a significant improvement in the model's detection performance after incorporating the DAFE module, with both AP and AR increasing by approximately 15\%. This once again demonstrates the effectiveness of the DAFE module, showcasing that a learning-based approach can outperform rule-based representation methods. Furthermore, the combination of PCF-Encoder and PCR modules also contributes to enhancing the model's performance to a certain extent. Although the metrics improvement it brings might not be as pronounced, the actual detection results demonstrate that incorporating phase information allows for a more accurate localization of targets.

\subsection{Interpreting Network with Grad-CAM++}
To analyze whether our network has indeed learned the correct features and to identify which features the network focuses on during the inference process, we employed GradCAM++ \cite{chattopadhay2018grad} for network visualization, as depicted in Fig. \ref{fig_13}. The image in the bottom-right corner overlays class activation maps (CAM) on the network's output results, visually indicating through heatmaps the areas in which our model detected the target. We can observe that the network accurately attends to the location of the target. In Fig. \ref{fig_13c} and \ref{fig_13d}, it's evident that besides focusing on the target itself, the network also pays attention to some spatial aspects nearby. This indicates that the network is learning and utilizing information from the surrounding noise to derive detection results.

\begin{figure}[!t]
\centering
\includegraphics[width=3.4in]{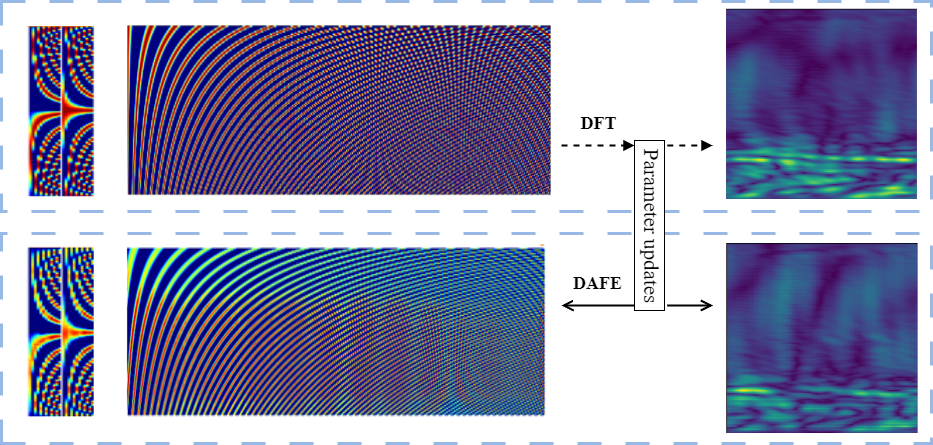}
\caption{The convolutional kernel parameters in DAFE change through training.}
\label{fig_14}
\end{figure}

Moreover, we present the evolution of DAFE convolutional kernel parameters during training in Fig. \ref{fig_14}. The upper section displays the initial stage of training, during which the kernel parameters align with the DFT. Through continuous parameter optimization and learning, the kernel parameters transition to the configuration shown in the lower section. At this point, the transformation is no longer a Fourier transform. We observe changes in the extracted feature heatmap, where a significant suppression of wall noise in the lower part of the heatmap occurs. The focus shifts more towards another segment containing weaker target signals amidst lower noise levels.

\section{Conclusion}
\label{Five}
In this paper, we propose TWP-CNN, a novel through-wall radar signal processing framework. The core module of this method is the feature extraction layer based on the DFT algorithm for raw ADC data. This layer, through further learning within the network beyond traditional algorithms, improves upon the drawbacks of traditional methods, namely their lack of adaptability and suboptimal performance. It significantly enhances the accuracy of human target detection in through-wall data with low signal-to-noise ratios caused by severe wall interference. Our self-collected through-wall radar dataset, RTW, containing raw ADC data, fills a void in this domain. We validate the effectiveness of our method on this dataset, achieving higher average precision and recall compared to other methods.

\bibliographystyle{IEEEtran}
\bibliography{conf}

\begin{thebibliography}{10}
\providecommand{\url}[1]{#1}
\csname url@samestyle\endcsname
\providecommand{\newblock}{\relax}
\providecommand{\bibinfo}[2]{#2}
\providecommand{\BIBentrySTDinterwordspacing}{\spaceskip=0pt\relax}
\providecommand{\BIBentryALTinterwordstretchfactor}{4}
\providecommand{\BIBentryALTinterwordspacing}{\spaceskip=\fontdimen2\font plus
\BIBentryALTinterwordstretchfactor\fontdimen3\font minus \fontdimen4\font\relax}
\providecommand{\BIBforeignlanguage}[2]{{%
\expandafter\ifx\csname l@#1\endcsname\relax
\typeout{** WARNING: IEEEtran.bst: No hyphenation pattern has been}%
\typeout{** loaded for the language `#1'. Using the pattern for}%
\typeout{** the default language instead.}%
\else
\language=\csname l@#1\endcsname
\fi
#2}}
\providecommand{\BIBdecl}{\relax}
\BIBdecl

\bibitem{zhao2023cubelearn}
P.~Zhao, C.~X. Lu, B.~Wang, N.~Trigoni, and A.~Markham, ``Cubelearn: End-to-end learning for human motion recognition from raw mmwave radar signals,'' \emph{IEEE Internet of Things Journal}, 2023.

\bibitem{yang2023adcnet}
B.~Yang, I.~Khatri, M.~Happold, and C.~Chen, ``{ADCNet: End-to-end perception with raw radar ADC data},'' \emph{arXiv preprint arXiv:2303.11420}, 2023.

\bibitem{kilicc2019through}
A.~K{\i}l{\i}{\c{c}}, {\.I}.~Babao{\u{g}}lu, A.~Babal{\i}k, A.~Arslan \emph{et~al.}, ``Through-wall radar classification of human posture using convolutional neural networks,'' \emph{International Journal of Antennas and Propagation}, vol. 2019, 2019.

\bibitem{nguyen2023high}
M.~Q. Nguyen, R.~Feger, T.~Wagner, and A.~Stelzer, ``High angular resolution method based on deep learning for fmcw mimo radar,'' \emph{IEEE Transactions on Microwave Theory and Techniques}, 2023.

\bibitem{ding2022human}
C.~Ding, L.~Zhang, H.~Chen, H.~Hong, X.~Zhu, and C.~Li, ``Human motion recognition with spatial-temporal-convlstm network using dynamic range-doppler frames based on portable fmcw radar,'' \emph{IEEE Transactions on Microwave Theory and Techniques}, vol.~70, no.~11, pp. 5029--5038, 2022.

\bibitem{ding2018application}
Y.~Ding, X.~Yu, J.~Zhang, and X.~Xu, ``Application of linear predictive coding and data fusion process for target tracking by doppler through-wall radar,'' \emph{IEEE Transactions on Microwave theory and Techniques}, vol.~67, no.~3, pp. 1244--1254, 2018.

\bibitem{song2021efficient}
Y.~Song, T.~Jin, Y.~Dai, and X.~Zhou, ``{Efficient through-wall human pose reconstruction using UWB MIMO radar},'' \emph{IEEE Antennas and Wireless Propagation Letters}, vol.~21, no.~3, pp. 571--575, 2021.

\bibitem{zheng2021human}
Z.~Zheng, J.~Pan, Z.~Ni, C.~Shi, S.~Ye, and G.~Fang, ``Human posture reconstruction for through-the-wall radar imaging using convolutional neural networks,'' \emph{IEEE Geoscience and Remote Sensing Letters}, vol.~19, pp. 1--5, 2021.

\bibitem{zheng2023radarformer}
Z.~Zheng, D.~Zhang, X.~Liang, X.~Liu, and G.~Fang, ``Radarformer: End-to-end human perception with through-wall radar and transformers,'' \emph{IEEE Transactions on Neural Networks and Learning Systems}, 2023.

\bibitem{li2020human}
H.~Li, G.~Cui, S.~Guo, L.~Kong, and X.~Yang, ``Human target detection based on {FCN} for through-the-wall radar imaging,'' \emph{IEEE geoscience and remote sensing letters}, vol.~18, no.~9, pp. 1565--1569, 2020.

\bibitem{pan2023multi}
J.~Pan, Z.~Zheng, D.~Zhao, K.~Yan, J.~Nie, B.~Zhou, and G.~Fang, ``A multi-target detection method based on improved {U-Net} for {UWB MIMO} through-wall radar,'' \emph{Remote Sensing}, vol.~15, no.~13, p. 3434, 2023.

\bibitem{wang2023real}
C.~Wang, D.~Zhu, L.~Sun, C.~Han, and J.~Guo, ``Real-time through-wall multi-human localization and behavior recognition based on mimo radar,'' \emph{IEEE Transactions on Geoscience and Remote Sensing}, 2023.

\bibitem{ye2019using}
W.~Ye, H.~Chen, and B.~Li, ``Using an end-to-end convolutional network on radar signal for human activity classification,'' \emph{IEEE Sensors Journal}, vol.~19, no.~24, pp. 12\,244--12\,252, 2019.

\bibitem{ye2020human}
W.~Ye and H.~Chen, ``Human activity classification based on micro-doppler signatures by multiscale and multitask fourier convolutional neural network,'' \emph{IEEE Sensors Journal}, vol.~20, no.~10, pp. 5473--5479, 2020.

\bibitem{ravanelli2018speaker}
M.~Ravanelli and Y.~Bengio, ``Speaker recognition from raw waveform with sincnet,'' in \emph{2018 IEEE spoken language technology workshop (SLT)}.\hskip 1em plus 0.5em minus 0.4em\relax IEEE, 2018, pp. 1021--1028.

\bibitem{stadelmayer2023parametric}
T.~Stadelmayer and A.~Santra, ``{Parametric convolutional neural network for radar-based human activity classification using raw ADC data},'' \emph{Authorea Preprints}, 2023.

\bibitem{stephan2021radar}
M.~Stephan, T.~Stadelmayer, A.~Santra, G.~Fischer, R.~Weigel, and F.~Lurz, ``{Radar image reconstruction from raw ADC data using parametric variational autoencoder with domain adaptation},'' in \emph{2020 25th International Conference on Pattern Recognition (ICPR)}.\hskip 1em plus 0.5em minus 0.4em\relax IEEE, 2021, pp. 9529--9536.

\bibitem{wang2021rodnet}
Y.~Wang, Z.~Jiang, Y.~Li, J.-N. Hwang, G.~Xing, and H.~Liu, ``Rodnet: A real-time radar object detection network cross-supervised by camera-radar fused object 3d localization,'' \emph{IEEE Journal of Selected Topics in Signal Processing}, vol.~15, no.~4, pp. 954--967, 2021.

\bibitem{caesar2020nuscenes}
H.~Caesar, V.~Bankiti, A.~H. Lang, S.~Vora, V.~E. Liong, Q.~Xu, A.~Krishnan, Y.~Pan, G.~Baldan, and O.~Beijbom, ``nuscenes: A multimodal dataset for autonomous driving,'' in \emph{Proceedings of the IEEE/CVF conference on computer vision and pattern recognition}, 2020, pp. 11\,621--11\,631.

\bibitem{ouaknine2021carrada}
A.~Ouaknine, A.~Newson, J.~Rebut, F.~Tupin, and P.~P{\'e}rez, ``Carrada dataset: Camera and automotive radar with range-angle-doppler annotations,'' in \emph{2020 25th International Conference on Pattern Recognition (ICPR)}.\hskip 1em plus 0.5em minus 0.4em\relax IEEE, 2021, pp. 5068--5075.

\bibitem{mostajabi2020high}
M.~Mostajabi, C.~M. Wang, D.~Ranjan, and G.~Hsyu, ``High-resolution radar dataset for semi-supervised learning of dynamic objects,'' in \emph{Proceedings of the IEEE/CVF Conference on Computer Vision and Pattern Recognition Workshops}, 2020, pp. 100--101.

\bibitem{wang2021rethinking}
Y.~Wang, G.~Wang, H.-M. Hsu, H.~Liu, and J.-N. Hwang, ``Rethinking of radar's role: A camera-radar dataset and systematic annotator via coordinate alignment,'' in \emph{Proceedings of the IEEE/CVF Conference on Computer Vision and Pattern Recognition}, 2021, pp. 2815--2824.

\bibitem{rebut2022raw}
J.~Rebut, A.~Ouaknine, W.~Malik, and P.~P{\'e}rez, ``Raw high-definition radar for multi-task learning,'' in \emph{Proceedings of the IEEE/CVF Conference on Computer Vision and Pattern Recognition}, 2022, pp. 17\,021--17\,030.

\bibitem{zhengliang2021dataset}
Z.~Zhengliang, Y.~Degui, Z.~Junchao, and T.~Feng, ``Dataset of human motion status using ir-uwb through-wall radar,'' \emph{Journal of Systems Engineering and Electronics}, vol.~32, no.~5, pp. 1083--1096, 2021.

\bibitem{tian2022uwb}
J.~Tian, S.~Yongkun, D.~Yongpeng, H.~Xikun, S.~Yongping, Z.~Xiaolong, and Q.~Zhifeng, ``Uwb-ha4d-1.0: An ultra-wideband radar human activity 4d imaging dataset,'' \emph{Journal of Radars}, vol.~11, no.~1, pp. 27--39, 2022.

\bibitem{cao2017realtime}
Z.~Cao, T.~Simon, S.-E. Wei, and Y.~Sheikh, ``Realtime multi-person 2d pose estimation using part affinity fields,'' in \emph{Proceedings of the IEEE conference on computer vision and pattern recognition}, 2017, pp. 7291--7299.

\bibitem{chattopadhay2018grad}
A.~Chattopadhay, A.~Sarkar, P.~Howlader, and V.~N. Balasubramanian, ``Grad-cam++: Generalized gradient-based visual explanations for deep convolutional networks,'' in \emph{2018 IEEE winter conference on applications of computer vision (WACV)}.\hskip 1em plus 0.5em minus 0.4em\relax IEEE, 2018, pp. 839--847.

\end{thebibliography}


\end{document}